%% file: paper.tex
\documentclass{article}

\usepackage[acronyms,nonumberlist,nopostdot,nomain,nogroupskip,acronymlists={hidden}]{glossaries}
\newglossary[algh]{hidden}{acrh}{acnh}{Hidden Acronyms}

\usepackage[preprint, nonatbib]{neurips_2023}

\usepackage[utf8]{inputenc} 
\usepackage[T1]{fontenc}    
\usepackage{hyperref}       
\usepackage{url}            
\usepackage{booktabs}       
\usepackage{amsfonts}       
\usepackage{nicefrac}       
\usepackage{microtype}      
\usepackage{graphicx} 
\usepackage{graphics}
\usepackage{caption} 
\usepackage{subfigure}
\usepackage{subcaption}
\usepackage{lscape}
\usepackage{float}
\usepackage{amssymb}
\usepackage{multirow}
\usepackage[capitalise]{cleveref}
\usepackage{tabularx}
\usepackage{enumitem}
\usepackage[table]{xcolor}
\usepackage{multirow}
\usepackage{appendix}

\title{Distributed Genetic Algorithm for Feature Selection}
  
\author{%
  M{\i}chael Potter \\
  Cognitive Systems Lab\\
  Center for SPIRAL \\
  Northeastern University, Boston, MA \\
  \texttt{potter.mi@northeastern.edu}
  \And
  Ayberk Yark{\i}n Y{\i}ld{\i}z \\
  Data, Networks, and Algorithms Lab\\
  Institute for the Wireless Internet of Things\\
  Northeastern University, Boston, MA  \\
  \texttt{yildiz.ay@northeastern.edu} \\
    \And
  Nishanth Marer Prabhu \\
  Northeastern University, Boston, MA  \\
  \texttt{marerprabhu.n@northeastern.edu} \\
    \And
  Cameron Gordon \\
  Northeastern University, Boston, MA  \\
  \texttt{gordon.ca@northeastern.edu} \\
}

\begin{document}
\input{acronyms}

\maketitle

\begin{abstract}
\label{abstract}%
We empirically show that process-based Parallelism speeds up the \gls{ga} for \gls{fs} 2x to 25x, while additionally increasing the \gls{ml} model performance on metrics such as F1-score, Accuracy, and \gls{roc-auc}. \\\textbf{Code:} \href{https://github.com/nishanthmarer/Genetic_Algorithm_Parallelize}{Github Repository}
\\\textbf{Video:} \href{https://drive.google.com/file/d/19I2xhl-S2vYEAGzsbOBFp_7hl8YAaF4Y/view?usp=sharing}{Final Project Video}
\end{abstract}

\section*{Notation}
\label{notation}%
\begin{table}[htbp]
    \small
    \centering
    \begin{tabular}{p{3cm}p{9cm}}
    \hline
    \textbf{Notation} & \textbf{Description} \\
    \hline
    $l$ & chromosome index \\
    $j$ & feature index \\
    $i$ & sample index \\
    $c_l$ & Chromosome \\
    $P$ & Population which consists of $L$ chromosomes \\
    $d$ & Number of features \\
    $c_{lj}$ & Gene (feature) $j$ of chromosome $l$ \\
    $s_l$ & Fitness score associated with chromosome $l$ \\
    $T_l$ & Time it takes \gls{ml} model $l$ to train on data subset corresponding to $c_l$ \\
    $N$ & Number of samples in a dataset \\
    $L$ & Number of chromosomes in a population. $L=|P|$ \\
    $x_i$ & Sample feature values for sample $i$ \\
    $y_i$ & Sample classification label for sample $i$ \\
    $f_j$ & a specific feature indexed by $j$ \\
    \hline
    \end{tabular}
    \caption{Notation Table}
    \label{tab:notation-table}
\end{table}
\vspace{-1cm}
\section*{Software}
\begin{itemize}
\item \textit{\textbf{\gls{slurm}}} \cite{osti_15002962}: An open-source job scheduler for Linux and Unix-like kernels on computer clusters. 
\item \textit{\textbf{sklearn}}\cite{scikit-learn}: An open-source Python library for machine learning featuring classification, regression, and clustering algorithms.
\item \textit{\textbf{PySpark}}\cite{pyspark}: Python API for Apache Spark, an open-source distributed computing system.
\item \textit{\textbf{joblib}}\cite{joblib}: Lightweight Python library to provide for embarrassingly parallel computations using multiprocessing.
\end{itemize}

\glsresetall
\section{Introduction}
\label{introduction}%
Many technology domains such as hyperspectral imagery, microarrays, and the internet  have created datasets with hundreds to hundreds of thousands of features \cite{tan2008genetic}. \gls{ml} tasks with high dimensionality pose challenges such as memory consumption, compute time, generalizability, and interpretability \cite{tan2008genetic}. Thus, selecting the most informative subset of features is desirable. However, the cardinality of all combinatorial feature subsets to search is $2^{d}$, which quickly becomes larger than the estimated number of atoms in the entire known universe for $d>270$.

\section{Methodology}

\subsection{Genetic Algorithm for Feature Selection} \label{GA_Algorithm_Explained}
The \gls{ga} for \gls{fs} is an optimization technique inspired by principles of natural selection and genetics. We use \gls{ga} to efficiently search through the large space of possible feature subsets to select the optimal subset of features. The primary components of the \gls{ga} are population size, crossover method, mutation rate, selection criteria, and evolution.

We denote a dataset as $\mathcal{D} = \{x_i,y_i\}_{i=1}^N$ = \{X,y\}, where $x_i \in \mathcal{R}^d$ is the sample feature values and $y_i \in \{0,1\}$ is the binary label for classification. We define the chromosome $c_l \in \mathcal{R}^d$ as a binary vector specifying which feature subset to use in \gls{ml} model training. The feature $j$ is incorporated into the feature subset if the gene $j$ in $c_l$ is expressed ($c_{lj}=1$); otherwise, we omit feature $j$ from the feature subset ($c_{lj}=0$). The data subset may be expressed as $\mathcal{D}_{ga}^{(l)} = \{X_{c_{lj}=1} ,y\}$. We show an example of how $c_l$ is used to convert $\mathcal{D}$ to $\mathcal{D}_{ga}^{(l)}$ in figure \ref{fig:example_ga_selection}:

\begin{figure}[h!]
    \centering
    \includegraphics[width=7cm]{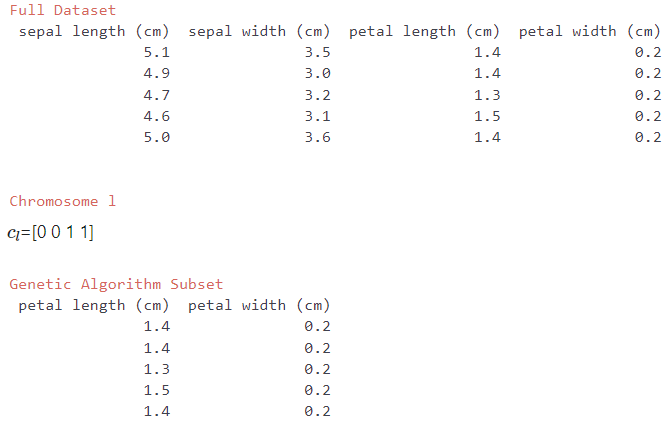}
    \caption{Example of chromosome being used for feature selection on the Iris Dataset \cite{bezdek1999will}}
    \label{fig:example_ga_selection}
\end{figure}

The fitness score $s_l$ associated to $c_l$ is the \gls{ml} model's performance (such as F1-score, accuracy or \gls{roc-auc}) on a validation split. 

\subsubsection{One-Point Cross-Over}
We employed a one-point crossover method to combine features from pairs of chromosomes. This method was chosen for its simplicity and effectiveness in maintaining the integrity of feature combinations.
The \gls{ga} one-point cross over operation between two chromosomes $c_1$ and $c_2$ will produce $c_3$ as:
\begin{align*}
    c_{3j} = \begin{cases}
    c_{1j} & \text{if } j \text{ is even} \\
    c_{2j} & \text{if } j \text{ is odd}
    \end{cases}
\end{align*}
, which produced a "child" chromosome that inherited genes from the pair of "parent" chromosomes. 

\subsubsection{Mutation}
Mutations of a population's chromosomes introduces variability for the next generation, which enables escaping local optima and exploring new regions of the feature space. 
The \gls{ga} mutation operation of chromosome $c_1$ may be expressed as 
\begin{align*}
    c_2 = c_1 \oplus [f_1, \ldots, f_d] \\
    f_j \sim Bernoulli(MR) 
\end{align*}
where $f_j = 1$ is a Bernoulli random variable with probability MR. When $f_j=1$, the gene expression of chromosome $c_{1j}$ is "flipped".

\subsubsection{Elitism}
The top $K$ chromosomes of a population, based on fitness scores, were retained in each generation (elitism), ensuring the preservation of high-quality feature sets for future generations. These chromosomes are untouched by \gls{ga} operators such as cross-over and mutation.

\subsubsection{Evolution Rounds}
Similar to the epoch number, number of evolution rounds considers the total number of generations for each run. The evolution of a population denotes the next generation after cross-over, mutation, and elitism operations. This evolution process is repeated until an user-specific convergence criteria is met, such as a metric threshold or a maximum number of evolution rounds.

\begin{figure}[h!]
\centering
\includegraphics[width=0.45\textwidth]{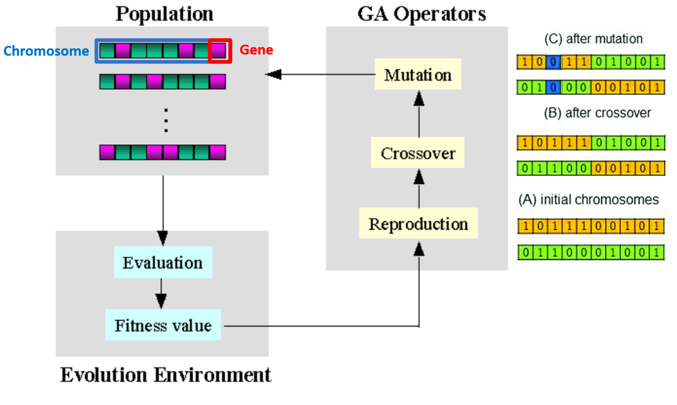}
\caption{General Genetic Algorithm Flow Diagram \cite{Liao_Sun_2001}}
\label{fig:GA-Diagram}
\end{figure}

A high level overview of the \gls{ga} is shown in Figure \ref{fig:GA-Diagram}.

\subsection{Parallelism of the Genetic Algorithm For Feature Selection}
\begin{figure}[h!]
\centering
\includegraphics[width=0.45\textwidth]{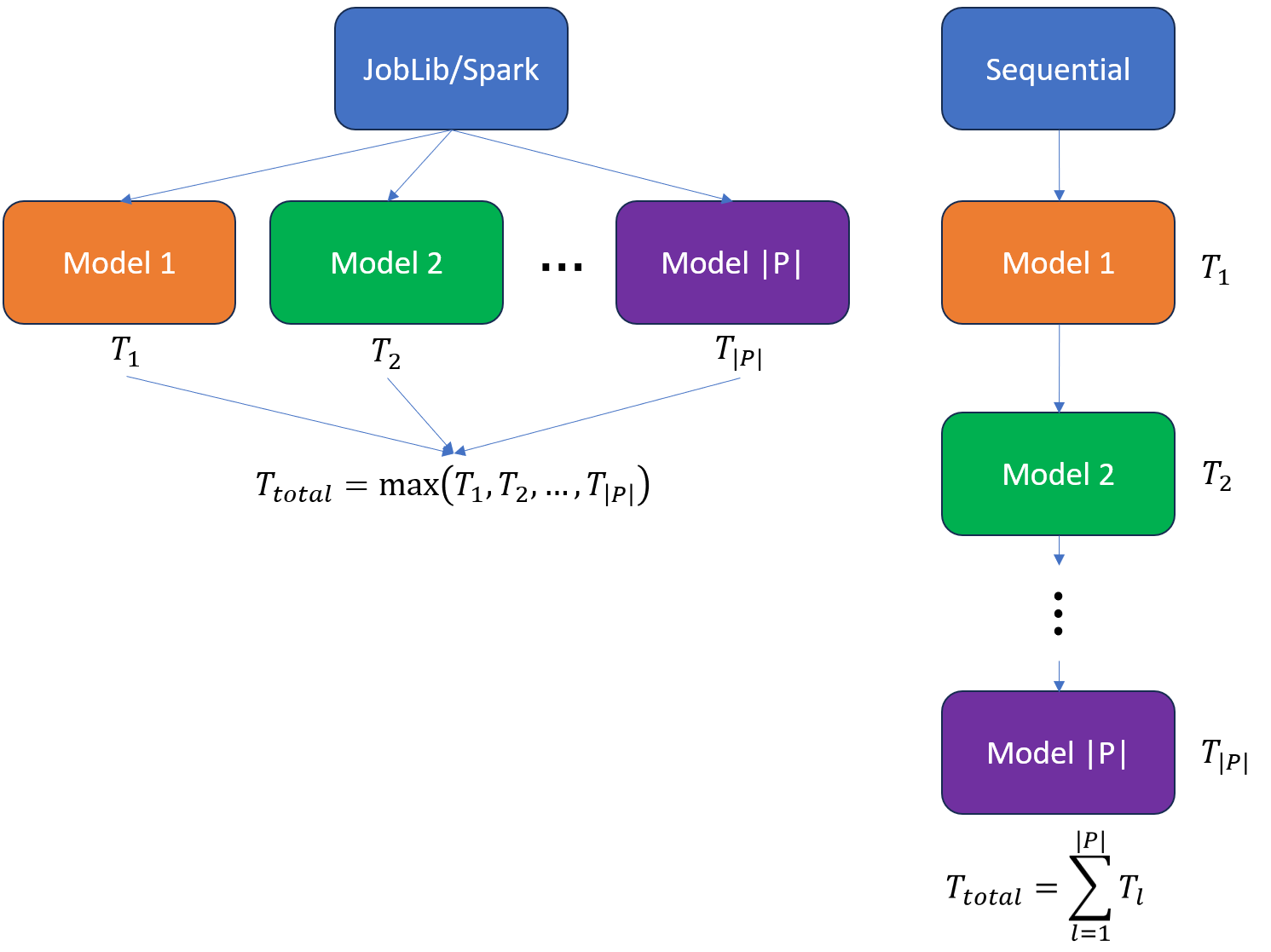}
\caption{Advantage in Time Consumption because of Parallelis. Figure only applies to one Evolution Round}
\label{fig:GAseqpar}
\end{figure}

We parallelize the evaluation of each chromosome's fitness score in the population $P=\{ c_l \}_{l=1}^{L}$. Normally, a \gls{ml} model is trained on each $\mathcal{D}^{(l)}_{ga}$, which results in training and validating as many \gls{ml} models as the number of chromosomes in the population. The sequential process is therefore prohibitively slow, as the total time of the \gls{ga} algorithm would be $\sum_{e=1}^{E} \sum_{l=1}^{L_e} T_l^{(e)}$. Therefore, parallelism will clearly decrease the time consumption of scoring a population at each evolution, as the total time of the \gls{ga} may be approximately $\sum_{e=1}^{E} \min (T_1^{(e)},T_2^{(e)},\ldots,T_{|P^{e}|}^{(e)})$ depending on the number of spawned processes (figure \ref{fig:GAseqpar}).

With a significant decrease in computation time, we may increase the number of evolution rounds and the population size to find a better optimal subset of features.

\subsection{Our Implementation}

\begin{figure}[h!]
\centering
\includegraphics[width=0.5\textwidth]{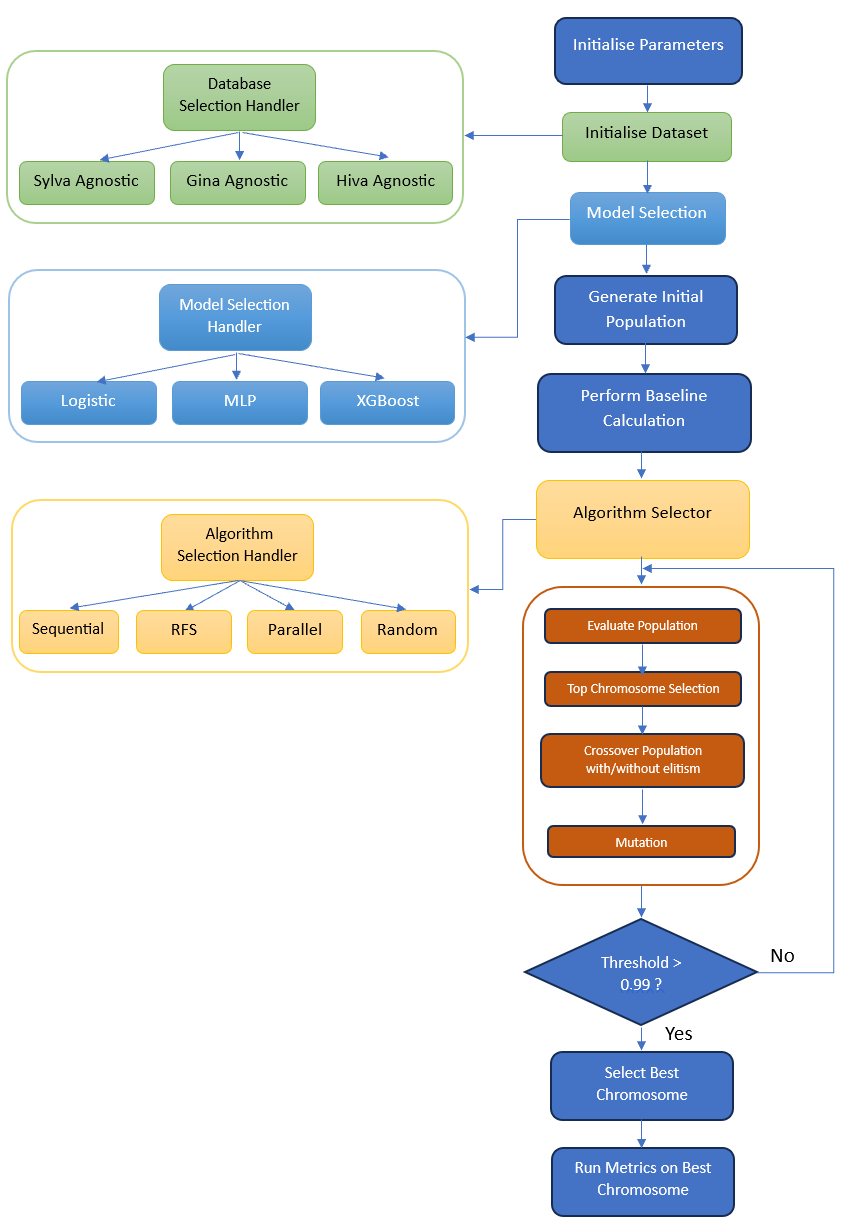}
\caption{Genetic Algorithm Code Flow}
\label{fig:GA-FlowChart}
\end{figure}

Let us look at the implementation, and refer to the Figure \ref{fig:GA-FlowChart}. The program starts initializing the parameters by loading the necessary libraries. Based on user input, the database selection handler loads the dataset. It performs pre-processing to obtain the features set and the target value, performs the required transformations, and splits the dataset into Training, Validation, and Testing sets in the configuration of $[70:20:10]$ and returns the result.

The model selection handler chooses from the variety of models available for fitting on the dataset and returns an object of the model.

To establish a baseline condition, an initial calculation is performed regarding the model’s performance on the complete dataset, and the resultant Accuracy, F1 score, and Execution time are recorded. 
The Algorithm Selector has the following choices: 
\begin{enumerate}[nolistsep]
\item \gls{ga} Sequential
\item Parallel (Spark and JobLib)
\item Random
\item \gls{rfs}
\end{enumerate}
In the case of the sequential process, the algorithm has two choices: utilize the Recursive Feature Selection \gls{rfs} or the \gls{ga} sequential algorithm. A sequential feature selector function is used in the former process, and the complete dataset is fit on the model. The output of the sequential selector is used to determine the best chromosomes (best features selected) on which the performance analysis is conducted. In the latter process, based on the population size, multiple combinations of chromosomes are applied to the dataset, and a collection of scores is obtained. Based on the explanation of the \gls{ga} mentioned in section \ref{GA_Algorithm_Explained}, the top performers are selected, and a new generation is obtained. This process continues until a threshold is satisfied or the number of iterations exceeds.

To optimize the runtime performance of the algorithm, Parallel execution provides flexibility to fit the $P$ models in parallel as depicted on the left side of the Figure \ref{fig:GAseqpar}.  In this process, two approaches have been examined. One uses Joblib, which provides a simple and easy parallel computing framework that enables the distribution of tasks on multiple threads or processes. Another approach is utilizing PySpark, an API for Apache Spark on Python. PySpark provides more granular controls regarding the number of partitions and executors, enabling the programmer to decide the extent of parallelism for the task.

In either of the above cases, the $P$ models are fit in parallel, and the resultant scores obtained are ranked, following which the \gls{ga} algorithm is applied for the best chromosome selection. This is repeated until the threshold is satisfied or the number of iterations exceeds the limit.

In the case of the random algorithm, a greedy approach is followed upon fitting $P$ populations on the model. The highest score and corresponding chromosome are considered the best feature combination selected. The following iterations do not utilize the best chromosome to optimize future generations. Instead, a new population is generated, and the process is repeated until the threshold is met or the number of iterations exceeds the given value.

Finally, the best chromosome obtained from the above algorithms is used for performance analysis. 

We show the best, worst, and average fitness score of a population over the number of evolution iterations for the \gls{mlp} with the \gls{ga-joblib} as an example in figure \ref{fig:GA-validation}:

\begin{figure}[h!]
\centering
\includegraphics[width=0.4\textwidth]{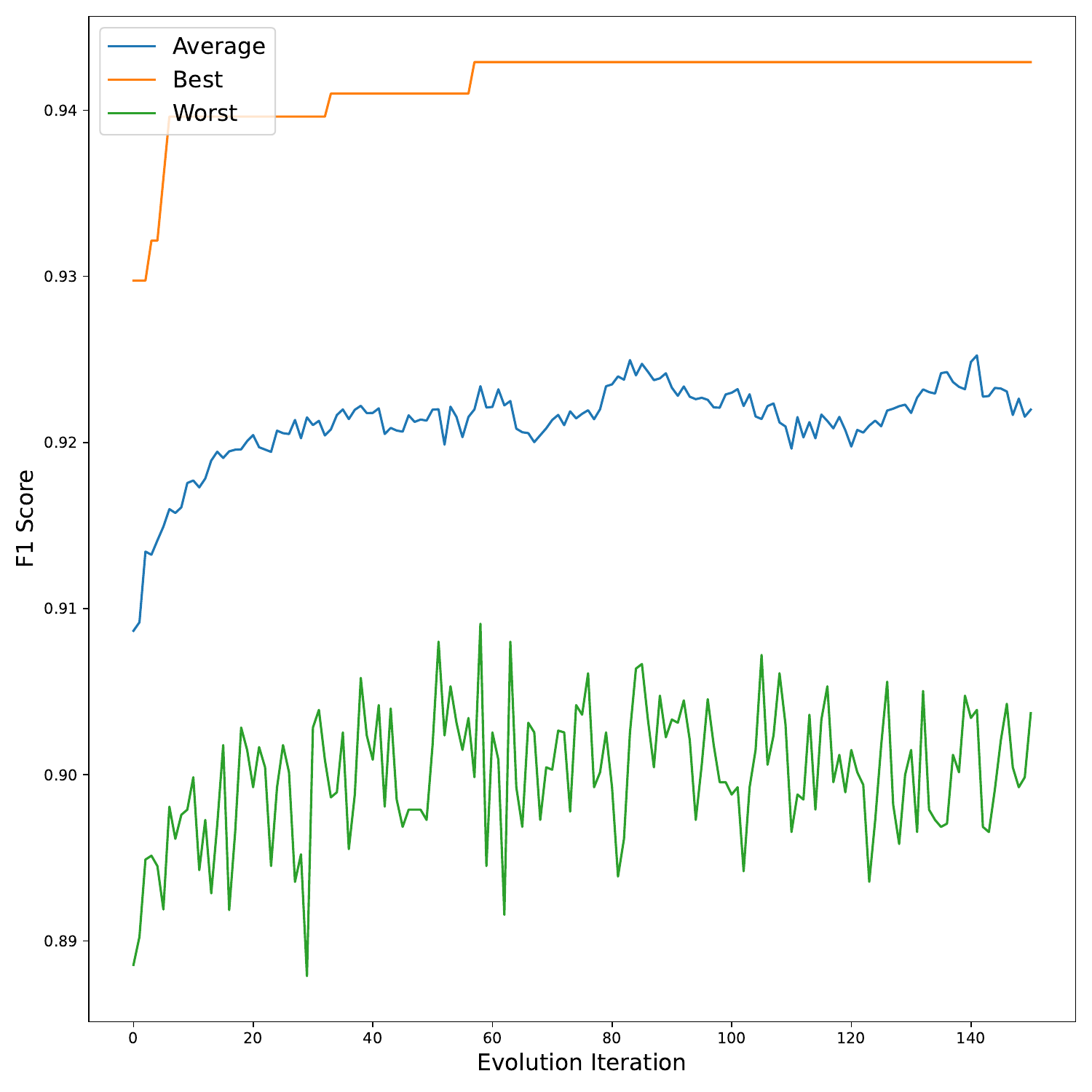}
\caption{Genetic Algorithm Validation Score per Iteration}
\label{fig:GA-validation}
\end{figure}

\section{Experiments}

We empirically show the speed up of parallelism for \gls{ga} for \gls{fs} with benchmarking combinations of models (\gls{mlp}, XGBoost, Logistic Regression), algorithms (\gls{rfs}, random, \gls{ga-seq}, \gls{ga-joblib}, \gls{ga-spark}) and datasets (Gina Agnostic, Silva Agnostic, and Hiva Agnostic).

\subsection{Configuration Details}
All experiments were run on Northeastern University Discover cluster by utilizing the \gls{slurm} job scheduler. The following sbatch configurations at Table \ref{tab:sbatch} were set for each experiment. We write a python script utilizing \textit{Popen} to submit and run all sbatch jobs in parallel. 

\begin{table}[h!]
    \centering
    \begin{tabular}{|p{2.5cm}|p{2.5cm}|p{2.5cm}|p{2.5cm}|p{2.5cm}|}
         \toprule
         \textbf{Partition} & \textbf{Memory} & \textbf{Cpus-per-task} &\textbf{Nodelist} & \textbf{Time} \\
         \midrule
         Short & 100Gb & 50 & d[] & 24:00:00 \\
         \bottomrule
    \end{tabular}
    \caption{SBATCH Configurations}
    \label{tab:sbatch}
\end{table}

Due to time constraints, we use the same set of hyperparameters for the \gls{ga} for \gls{fs} in every experiment (table \ref{tab:hyper}). We employed multiple \gls{ml} models, such as \gls{mlp}, XGBoost, and logistic regression, within each parallelism approach to assess the trade off between parallelism overhead and timing advantages.

\begin{table}[h!]
    \centering
    \begin{tabular}{|p{3.0cm}|p{3.0cm}|}
         \toprule
         \textbf{Hyper parameter} & \textbf{Value}\\
         \midrule
         mutation rate & 0.2\\
        \hline
         crossover method & one-point\\
        \hline
         elitism & 2\\
        \hline
         selection criteria & F1-Score\\
        \hline
         evolution rounds & 150 \\
        \hline
        population size & 150 \\
                \hline
        test set size & 0.1 \\
                \hline
        validation set size & 0.2 \\
         \bottomrule
    \end{tabular}
    \caption{Hyperparameter Settings for \gls{ga} for \gls{fs}}
    \label{tab:hyper}
\end{table}

\subsection{Datasets}
\label{data_descr}
We evaluated our approach on three different agnostic datasets from the \textit{Agnostic Learning vs Prior Knowledge} challenge \cite{guyon2007agnostic,OpenML2013}, which purposely consists of unnecessary features. Half of the features in the agnostic datasets are "distractors", and lead to degradation of classification metrics. Therefore, these datasets showcase the benefits of \gls{fs}. The challenge was to see if task performance on domain-specific feature engineered datasets are matched by \gls{ml} models trained on datasets without any domain-specific knowledge (agnostic).  \cite{OpenML2013}.

\begin{table}[!h]
\begin{center}
\begin{tabular}{|c | c c c|} 
\toprule
Dataset & Dimension ($d$) & Samples ($N)$ & Classes \\ [0.5ex] 
\midrule
Sylva Agnostic & 217 & 14395 & 2 \\
Gina Agnostic & 451 & 174 & 2 \\
Hiva Agnostic & 1203 & 171 & 2 \\ 
\bottomrule
\end{tabular}
\caption{Dataset Specifications \cite{OpenML2013}}
\label{table:data_table}
\end{center}
\end{table}

The datasets we analyze (table \ref{table:data_table}) are described below.

\subsubsection{Sylva Agnostic}
Classify 30x30 meter forest cover images as Ponderosa pine or other. Each sample is 4 records, where 2 true records match the target and 2 records are random.

\subsubsection{Gina Agnostic}
Classify digits as even or odd  from an MNIST subset. Each sample is a two digit number, where only the unit digit is informative for the task.

\subsubsection{Hiva Agnostic}
Classify compounds against AIDS HIV infection as active or inactive.

\subsection{Method of Parallelism}
\subsubsection{PySpark}

PySpark is an API for Apache Spark on Python. We have utilized pyspark to enable the parallelization of tasks, which enables data processing much faster. From Figure \ref{fig:GAseqpar}, we can see how spark helps in parallelization compared to the sequential method. Spark allows the developer to control the extent of parallelism and partitioning of the dataset across multiple machines.

For our use case, we have demonstrated the use of pyspark to fit $P$ models which contain different chromosome combinations in parallel. By performing the model fitting in parallel, the total time taken for execution depends on the slowest execution process.

\subsubsection{JobLib}
Joblib is a framework which provides an embarrassingly parallel computing framework that enables task distribution across multiple machines or within a single machine. JobLib provdies flexibity to the user to change between multiple backends. These backends include multithreading, multiprocessing, usage of open source parallelization frameworks such as Ray and Dask.

For our use case, we explored the usage of multithreading and instructed joblib to utilise multiple cores from a CPU and fit $P$ models in parallel.

\section{Results}

We show that the computational time of the \gls{ga} for \gls{fs} is reduced by a factor of 2x to 25x (table \ref{table:runtime}. The larger reduction in computational time correlates with the complexity of the \gls{ml} model. Furthermore, the Accuracy, F1, and ROC-AUC increased from the baseline for most of the dataset and model combinations when using the \gls{ga} for \gls{fs} (table \ref{table:results}). 

\subsection{Computation Time}
From the perspective of parallelism or time efficiency, the \gls{ga-joblib} algorithm outperformed the \gls{ga-spark} algorithm in the context of the MLP and logistic regression models. Notably, the \gls{ga-seq} method demonstrated a significant increase in runtime, which underscores the benefits of parallel processing in \gls{fs} tasks. The random approach varied across datasets, highlighting the unpredictability and inefficiency of non-guided \gls{fs} methods. We note the baseline metrics took the least time, emphasizing the total computational cost of the \gls{fs} process.

\begin{table}[h!]
\centering
\begin{tabular}{lllll}
\toprule
 & dataset & GINA AGNOSTIC & HIVA AGNOSTIC & SYLVA AGNOSTIC \\
model & algorithm &  &  &  \\
\midrule
\multirow[t]{6}{*}{mlp} & GA SPARK & 17.541 & 46.471 & 20.441 \\
 & GA JOBLIB & 13.558 & 35.05 & 20.806 \\
 & RANDOM & 280.24 & 503.329 & 493.163 \\
 & GA SEQ & 250.116 & 494.79 & 535.266 \\
 & RFS & nan & nan & nan \\
 & BASELINE & 2.259 & 5.301 & 3.589 \\
\cline{1-5}
\multirow[t]{6}{*}{xgboost} & GA SPARK & 43.835 & 31.333 & 24.76 \\
 & GA JOBLIB & 22.16 & 34.257 & 25.515 \\
 & RANDOM & 322.759 & 72.523 & 227.809 \\
 & GA SEQ & 90.887 & 72.568 & 66.404 \\
 & RFS & nan & nan & nan \\
 & BASELINE & 0.872 & 1.177 & 0.869 \\
\cline{1-5}
\multirow[t]{6}{*}{logistic} & GA SPARK & 5.768 & 5.336 & 2.601 \\
 & GA JOBLIB & 9.455 & 9.166 & 2.775 \\
 & RANDOM & 12.028 & 11.918 & 19.987 \\
 & GA SEQ & 7.429 & 11.679 & 22.798 \\
 & RFS & 44954.382 & nan & 16359.015 \\
 & BASELINE & 0.117 & 0.186 & 0.37 \\
\cline{1-5}
\bottomrule
\end{tabular}
\caption{\centering Average Running Times (seconds) per Generation of \gls{fs}. The baseline is the model fit on all the features once. }
\label{table:runtime}
\end{table}

\subsection{Model Performance}
Regarding model performance, the \gls{ga-spark} and \gls{ga-joblib} methods with the XGBoost model yielded high scores across all metrics, indicating the effectiveness of the \gls{fs} process when combined with a powerful ensemble model. The MLP and logistic models also showed improved performance with \gls{fs} , albeit with some variations across metrics and datasets. It is evident that genetic algorithms with parallelism not only accelerated the process but also maintained, and in some cases enhanced, the model performance. 


\begin{table}[h!]
    \centering
    \scalebox{0.8}{
\begin{tabular}{lllllllllll}
\toprule
 & dataset & \multicolumn{3}{c}{GINA AGNOSTIC} & \multicolumn{3}{c}{HIVA AGNOSTIC} & \multicolumn{3}{c}{SYLVA AGNOSTIC} \\
 &  & Accuracy & F1 & ROC AUC & Accuracy & F1 & ROC AUC & Accuracy & F1 & ROC AUC \\
model & algorithm &  &  &  &  &  &  &  &  &  \\
\midrule
\multirow[t]{6}{*}{mlp} & GA SPARK & 0.89 & 0.888 & 0.89 & 0.972 & 0.5 & 0.696 & 0.994 & 0.95 & 0.976 \\
 & GA JOBLIB & 0.891 & 0.889 & 0.891 & 0.972 & 0.5 & 0.696 & 0.994 & 0.95 & 0.976 \\
 & RANDOM & 0.879 & 0.876 & 0.879 & 0.972 & 0.455 & 0.664 & 0.991 & 0.929 & 0.974 \\
 & GA SEQ & 0.893 & 0.891 & 0.893 & 0.972 & 0.5 & 0.696 & 0.994 & 0.95 & 0.976 \\
 & RFS & NaN & NaN & NaN & NaN & NaN & NaN & NaN & NaN & NaN \\
 & BASELINE & 0.879 & 0.875 & 0.879 & 0.969 & 0.381 & 0.631 & 0.993 & 0.944 & 0.97 \\
\cline{1-11}
\multirow[t]{6}{*}{xgboost} & GA SPARK & 0.925 & 0.924 & 0.925 & 0.965 & 0.286 & 0.596 & 0.992 & 0.938 & 0.964 \\
 & GA JOBLIB & 0.925 & 0.924 & 0.925 & 0.967 & 0.364 & 0.63 & 0.994 & 0.955 & 0.976 \\
 & RANDOM & 0.931 & 0.931 & 0.931 & 0.967 & 0.3 & 0.598 & 0.992 & 0.938 & 0.964 \\
 & GA SEQ & 0.925 & 0.924 & 0.925 & 0.967 & 0.364 & 0.63 & 0.994 & 0.955 & 0.976 \\
 & RFS & nan & nan & nan & nan & nan & nan & nan & nan & nan \\
 & BASELINE & 0.931 & 0.928 & 0.93 & 0.969 & 0.316 & 0.599 & 0.992 & 0.932 & 0.959 \\
\cline{1-11}
\multirow[t]{6}{*}{logistic} & GA SPARK & 0.804 & 0.795 & 0.804 & 0.962 & 0.429 & 0.691 & 0.992 & 0.931 & 0.954 \\
 & GA JOBLIB & 0.805 & 0.8 & 0.805 & 0.966 & 0.432 & 0.677 & 0.992 & 0.931 & 0.954 \\
 & RANDOM & 0.795 & 0.793 & 0.795 & 0.962 & 0.385 & 0.659 & 0.994 & 0.949 & 0.97 \\
 & GA SEQ & 0.804 & 0.795 & 0.804 & 0.967 & 0.462 & 0.694 & 0.993 & 0.944 & 0.97 \\
 & RFS & 0.801 & 0.801 & 0.801 & nan & nan & nan & 0.988 & 0.895 & 0.93 \\
 & BASELINE & 0.798 & 0.797 & 0.798 & 0.965 & 0.4 & 0.661 & 0.993 & 0.944 & 0.97 \\
\cline{1-11}
\bottomrule
\end{tabular}}
\caption{\centering Test Results over Different Metrics. The baseline is the model fit on all the features once.}
\label{table:results}
\end{table}

\subsection{Selected Features}
We compute the Jaccard Overlap as a measure of similarity between the selected feature subsets (best chromosomes) between algorithms to observe the differences in feature selection between \gls{ga} for \gls{fs}, \gls{rfs}, and random methods:

\begin{align*}
    J(c_1,c_2) = \frac{|c_1 \cap c_2|}{|c_1 \cup c_2|} \in [0,1]
\end{align*}

\begin{figure}[h!]
    \centering
    \includegraphics[width=14cm]{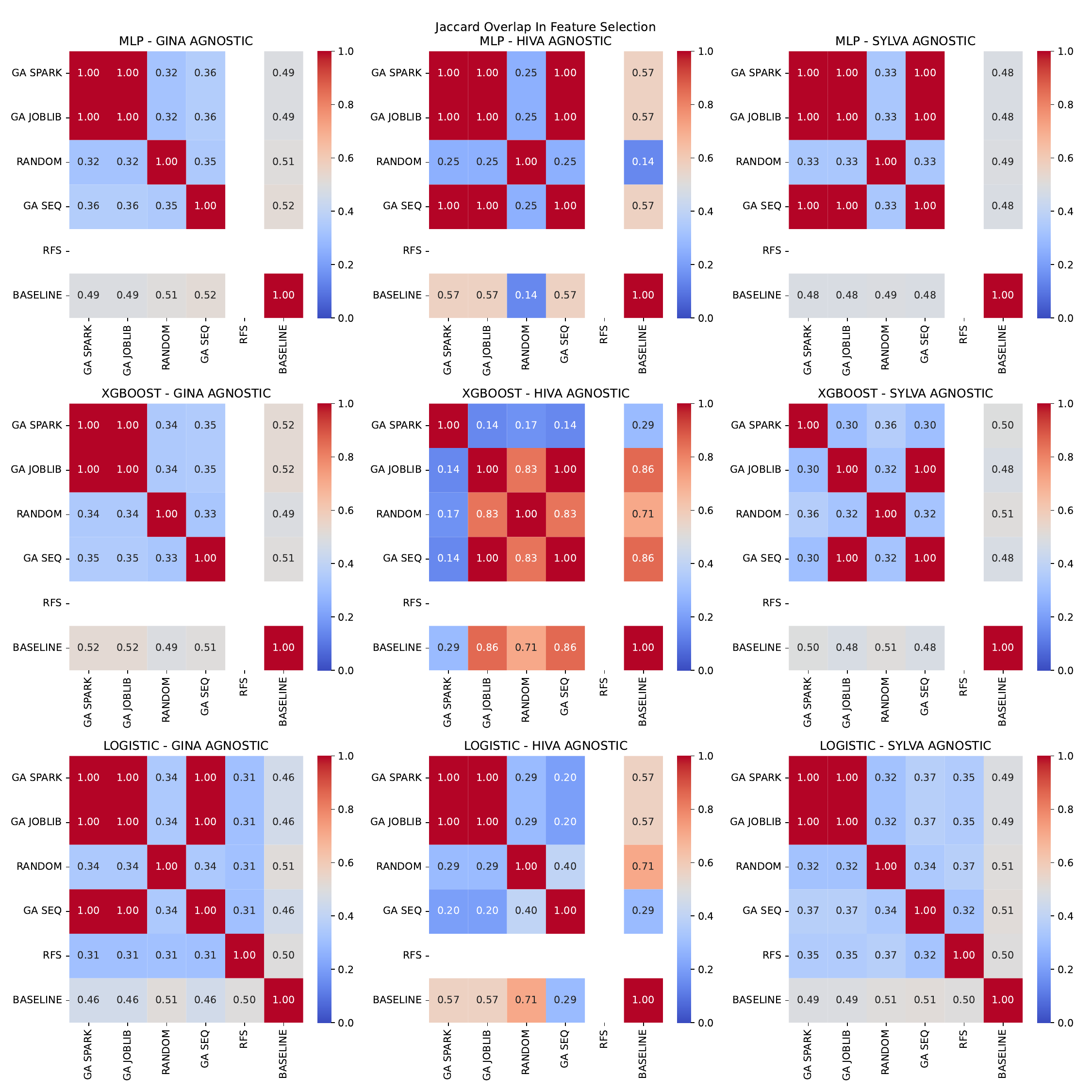}
    \caption{The Jaccard Overlap of the Best Chromosome Between Algorithms. The complete white means that there is NaN for one of the methods (\gls{rfs} in this case timed out on the short partition 24H time limit)}
    \label{fig:featureoverlap}
\end{figure}

The higher the Jaccard Overlap, the higher the overlap in selected features is between two methods. We note that the Jaccard Overlap values of $\approx 0.5$ in the row for baseline method in figure \ref{fig:featureoverlap} is in alignment with the our expectations. The Jaccard Overlaps are $\approx 0.5$ because the datasets in section \ref{data_descr} are known to have exactly half disctractor features, which do not improve predictive performance. However, we note that the random and \gls{rfs} only share $\approx=0.33$ of the same features selected with the \gls{ga} for \gls{fs} methods. 

\section{Reproducibility}
\label{reproducibility}
We run \gls{ga} for \gls{fs} with three different parallelism algorithms denoted \gls{ga-spark}, \gls{ga-joblib}, and \gls{ga-seq}. The underlying \gls{ga} computation does not change with respect to the parallelism algorithms, thus we expect the same deterministic results if the random seeds are set properly on pseudo-random number generators. However, we were unable to set the random seeds properly as we saw different feature subsets selected based on the method choice (sometimes the same feature subset was selected, sometimes not). For example, XGBoost may still exhibit non-deterministic results due to non-determinism in floating point summation order and multi-threading \cite{Chen:2016:XST:2939672.2939785}. Ultimately, there was an underlying issue with reproducibility and multiprocesses that we were unable to address, but hope to in the future. It should be noted that the random seeds worked on a local computer with Windows operating system (Appendix \ref{Appendix}). We do not feel that this issue detracts from the validity of our experiments with respect to timing and predictive improvements.

\section{Conclusion}
\label{conclusion}

This work presented a comprehensive study on the application of a genetic algorithm (GA) for feature selection (FS) in machine learning (ML) tasks with high-dimensional data. Through empirical evidence, we demonstrated that incorporating parallelism within the \gls{ga} significantly enhances computational efficiency, with a 2x to 25x speedup, depending on the ML model and dataset used. The convergence of parallel computing techniques with genetic algorithms has proven not only to reduce the time complexity of the FS process but also to maintain or even improve the performance of ML models in terms of F1-score, accuracy, and ROC-AUC metrics. The parallelized \gls{ga} successfully identified optimal feature subsets that enhanced model performance, reinforcing the necessity and effectiveness of \\gls{fs} in ML. The integration of parallel processing frameworks like Joblib and PySpark showcased the feasibility of applying \gls{ga} for FS on a larger scale, catering to the needs of modern data-intensive applications.

In conclusion, the \gls{ga} with parallel processing capabilities significantly reduces the time required for \gls{fs} in extremely high dimensional datasets while preserving or improving the predictive performance of \gls{ml} models. This emphasizes the utility of parallelism in handling high-dimensional data in \gls{ml} tasks.

\clearpage

\include{Appendix}

\clearpage
\bibliography{paper}

\end{document}

%% file: acronyms.tex
\newacronym{ga}{GA}{Genetic Algorithm}
\newacronym{fs}{FS}{Feature Selection}
\newacronym{ml}{ML}{Machine Learning}
\newacronym{slurm}{SLURM}{Simple Linux Utility for Resource Management}
\newacronym{roc-auc}{ROC-AUC}{Receiver Operating Characteristic Area Under the Curve}
\newacronym{fs}{FS}{Feature Selection}
\newacronym{mlp}{MLP}{Multilayer Perceptron}
\newacronym{rfs}{RFS}{Recursive Feature Selection}
\newacronym{ga-seq}{GA-Sequential}{Genetic Algorithm - Sequential}
\newacronym{ga-spark}{GA-Spark}{Genetic Algorithm - Spark}
\newacronym{ga-joblib}{GA-Joblib}{Genetic Algorithm - Joblib}

%% file: Appendix.tex
\appendix

\begin{appendices}
\section{Appendix A}
\label{Appendix}
\begin{figure}[h!]
    \centering
    \includegraphics[width=11cm]{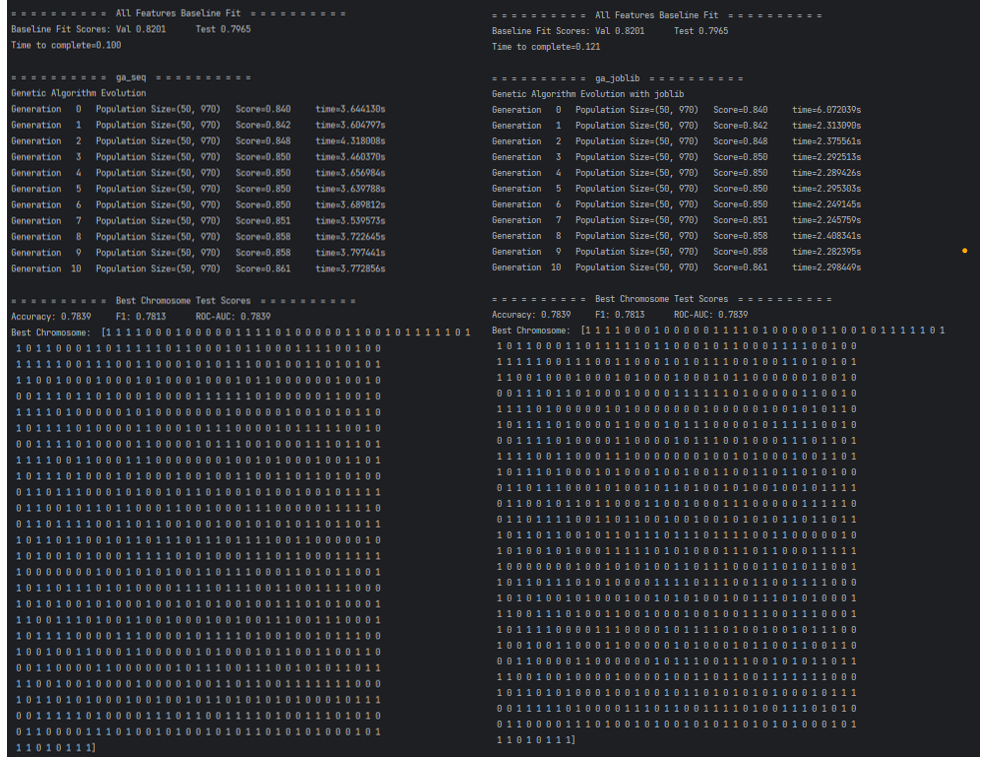}
    \caption{Logistic Regression on the GINA Agnostic dataset with GA-SEQ and GA-JOBLIB}
    \label{fig:reproduce}
\end{figure}
\end{appendices}

%% file: paper.bbl
\begin{thebibliography}{10}

\bibitem{osti_15002962}
M.~Jette, C.~Dunlap, J.~Garlick, and M.~Grondona, ``Slurm: Simple linux utility for resource management,''

\bibitem{scikit-learn}
F.~Pedregosa, G.~Varoquaux, A.~Gramfort, V.~Michel, B.~Thirion, O.~Grisel, M.~Blondel, P.~Prettenhofer, R.~Weiss, V.~Dubourg, J.~Vanderplas, A.~Passos, D.~Cournapeau, M.~Brucher, M.~Perrot, and E.~Duchesnay, ``Scikit-learn: Machine learning in {P}ython,'' {\em Journal of Machine Learning Research}, vol.~12, pp.~2825--2830, 2011.

\bibitem{pyspark}
{PySpark Development Team}, ``Pyspark is the python api for apache spark,'' 2009.

\bibitem{joblib}
{Joblib Development Team}, ``Joblib: running python functions as pipeline jobs,'' 2020.

\bibitem{tan2008genetic}
F.~Tan, X.~Fu, Y.~Zhang, and A.~G. Bourgeois, ``A genetic algorithm-based method for feature subset selection,'' {\em Soft Computing}, vol.~12, pp.~111--120, 2008.

\bibitem{bezdek1999will}
J.~C. Bezdek, J.~M. Keller, R.~Krishnapuram, L.~I. Kuncheva, and N.~R. Pal, ``Will the real iris data please stand up?,'' {\em IEEE Transactions on Fuzzy Systems}, vol.~7, no.~3, pp.~368--369, 1999.

\bibitem{Liao_Sun_2001}
Y.-H. Liao and C.-T. Sun, 2001.

\bibitem{guyon2007agnostic}
I.~Guyon, A.~Saffari, G.~Dror, and G.~Cawley, ``Agnostic learning vs. prior knowledge challenge,'' in {\em 2007 International Joint Conference on Neural Networks}, pp.~829--834, IEEE, 2007.

\bibitem{OpenML2013}
J.~Vanschoren, J.~N. van Rijn, B.~Bischl, and L.~Torgo, ``Openml: Networked science in machine learning,'' {\em SIGKDD Explorations}, vol.~15, no.~2, pp.~49--60, 2013.

\bibitem{Chen:2016:XST:2939672.2939785}
T.~Chen and C.~Guestrin, ``{XGBoost}: A scalable tree boosting system,'' in {\em Proceedings of the 22nd ACM SIGKDD International Conference on Knowledge Discovery and Data Mining}, KDD '16, (New York, NY, USA), pp.~785--794, ACM, 2016.

\end{thebibliography}
